\documentstyle[preprint,aps]{revtex}
\draft

\begin{document}
\title{Quasi 1 and 2d Dilute Bose Gas in Magnetic Traps : Existence of
Off-Diagonal
Order and Anomalous Quantum Fluctuations}
\author{Tin-Lun Ho$^{+}$ and Michael Ma$^{\ast}$}
\address{$^{+}$Physics Department, The Ohio State University, Columbus,
Ohio 43210 \\
$^{\ast}$Physics Department, University of Cincinnati, Cincinnati, Ohio 45221}
\maketitle

\begin{abstract}
Current magnetic traps can be made so anisotropic that the atomic gas
trapped inside behaves like quasi one or two dimensional system. Unlike the
homogeneous case, quantum phase fluctuations do not destroy macroscopic
off-diagonal order of trapped Bose gases in $d\leq 2$. In the dilute limit,
quantum fluctuations increase, remain constant, and decrease with size for $%
3, 2, 1d$ respectively. These behaviors are due to the combination of a
finite gap and the universal spectrum of the collective mode.
\end{abstract}

\vspace{0.2in}


\vspace{0.2in}

The technology of magnetic traps has advanced rapidly since the first series
of reports of Bose-Einstein condensation in atomic gases of $^{87}$Rb\cite
{Rb}, $^{23}$Na\cite{Na}, and $^{7}$Li\cite{Li}. The latest generation of
magnetic traps\cite{cloverleaf}\cite{Coloradomix} is capable of trapping $%
10^{6}$ atoms in the ground state, three orders of magnitudes higher than
that in the first Rb condensate experiment\cite{Rb}. A common characteristic
of these traps is that they are very anisotropic. The cylindrically
symmetric harmonic potentials in these traps have transverse frequency $%
\omega_{\perp}$ about 20 times larger than the longitudinal frequency $%
\omega_{z}$. The resulting condensate looks like a cigar aligned along the
symmetry axis $z$. With current technology, there appears no difficulty in
achieving even higher asymmetry, say, with $\omega_{\perp}/\omega_{z}\sim
100 $. In such limit, one can produce atomic gases with all the atoms lying
in the lowest harmonic oscillator state in the $xy$-plane, leaving the
motion along $z$ the only degrees of freedom. The system then behaves like a
one dimensional ($1d$) Bose gas\cite{Ketterle1d}. Similarly, for $\omega
_{\perp }/\omega _{z}<<1$, the system behaves like a $2d$ system\cite{Ber}.

The possibility of realizing these low dimensional systems raises the
question of the effects of quantum fluctuations. For homogeneous Bose
systems, it is well known that phase fluctuations destroy finite (zero)
temperature Bose condensation in dimension $d\leq 2(\leq 1)$. The situation,
however, is different for trapped Bose gases. For non-interacting bosons,
finite spacing between the lowest and the next energy level allow BEC to
occur at finite temperatures even in 1D \cite{Ketterle1d} \cite{comment}.
The absence of gapless excitations also means that BEC will not be
destroyed immediately as interactions between bosons are turned on.
Nevertheless, for trapped bosons, quantum fluctuation effects due to
interactions do show up in an unexpected way, which is the subject of this
paper. We shall show that while the gapped nature of the collective modes
eliminates the usual
infrared divergence in all dimensions, the number of low energy modes
contributing significantly to quantum fluctuation increases with the system
size. As a result, quantum fluctuations suppresses but do not destroy the
condensate. They also grow with size at a rate that increases with
dimensionality.  More precisely, we find that

\noindent {\bf (I)} The density matrix $W({\bf r},{\bf r^{\prime }}) $ of a $%
d $-dimensional dilute Bose gas in a harmonic trap exhibits macroscopic
off-diagonal order of the form $W({\bf r},{\bf r^{\prime }})$$\rightarrow $ $%
\phi ^{*}({\bf r})\phi ({\bf r^{\prime }})$ for $|{\bf r}-{\bf r^{\prime }}|$
of the order of the system size $R$, (both ${\bf r}$ and ${\bf r^{\prime }}$
are inside the atom cloud); $\rho _{o}$ is the density of the system, $\phi (%
{\bf r})$$=$$\sqrt{\rho _{o}({\bf r})}e^{-\gamma _{d}/2}$ (up to a phase
factor), and $\gamma _{d}$ is a constant arising from the Gaussian
fluctuations about the mean field solution : $\gamma _{1}\propto R^{-1}{\rm %
ln}(R/\sigma )$, $\gamma _{d}\propto (R/\sigma )^{d-2}$ for $d\geq 2$, where
$\sigma $ is the width of the ground state Gaussian of the $d$ -dimensional
harmonic well with frequency $\omega $, $\sigma =\sqrt{\hbar /M\omega }$.
These results show that {\em within the dilute regime}, the Gaussian
approximation becomes more (less) accurate with increasing size in $1d$ ($3d$%
). This anomalous demensionality dependence of the quantum fluctuations can
be directly confirmed experimentally by studying how the momentum
distribution depends on the number of particles.

\noindent {\bf (II)} The behavior in {\bf (I)} is due to the gapped nature
as well as the {\em universal} feature of the collective mode. For $d=1$,
the collective modes are $\Omega _{\ell }=\omega \sqrt{\ell (\ell +1)/2}$, $%
\ell =1,2,..$. For $d=2$, they are $\Omega _{n,m}=\omega \sqrt{%
(n^{2}-m^{2}+2n)/2}$, $n=|m|+2p$, $p=0,1,2,..$, $m=$integer.

We shall first derive these results, and then discuss their implications for
real atomic gases. Let us consider a $d$-dimensional dilute Bose gas with a
partition function $Z={\rm Tr}e^{-\beta (H-\mu N)}$, where $\mu $ is the
chemical potential, $K=H-\mu N$$=\int {\rm d}^{d}{\bf r}$$[(\hbar ^{2}/2M)$ $%
{\bf \nabla }\hat{\psi}^{+}\cdot {\bf \nabla }\hat{\psi}$$+(U-\mu )$ $\hat{%
\psi}^{+}\hat{\psi}$ $+(g_{d}/2)\hat{\psi}^{+}\hat{\psi}^{+}\hat{\psi}\hat{%
\psi}]$, $g_{d}>0$, $U({\bf r})=(1/2)M\omega ^{2}|{\bf r}|^{2}$ is a $d$
-dimensional harmonic well, and ${\bf \nabla }$ is a $d$-dimensional
gradient. $H$ is an effective low energy Hamiltonian where the Fourier
component of $\hat{\psi}$ with wavelength shorter than an atomic scale $%
a_{c} $ has been coarse-grained out.

It is convenient to write the partition function as a coherent state path
integral ${\cal Z}=\int D\phi ^{*}D\phi e^{-S/\hbar }$, where
\begin{equation}
S=\int \left[ \hbar \phi ^{*}\partial _{\tau }\phi +\frac{\hbar ^{2}}{2M}|%
{\bf \nabla }\phi |^{2}+(U-\mu )|\phi |^{2}+\frac{g_{d}}{2}|\phi
|^{4}\right] ,  \label{S}
\end{equation}
with $\int \equiv \int {\rm d}^{d}{\bf r}\int_{0}^{\beta }{\rm d}\tau $.
Anticipating a condensate, we write $\phi =e^{i\theta }\sqrt{\rho }$
. The stationary
point is a constant phase $\theta _{o}$ and a density $\rho _{o}({\bf r})$
satisfying the Gross-Piteavksii (GP) equation,
\begin{equation}
-\frac{\hbar ^{2}}{2M}{\bf \nabla }^{2}\sqrt{\rho _{o}}+(U-\mu )\sqrt{\rho
_{o}}+g_{d}\rho _{o}\sqrt{\rho _{o}}=0.  \label{GP}
\end{equation}

Expanding $S$ about $(\theta _{o},\rho _{o}({\bf r}))$ to quadratic order in
phase fluctuations $\theta _{1}$ and density fluctuations $\rho _{1}$, $%
(\rho _{1}=\rho -\rho _{o},\theta _{1}=\theta -\theta _{o})$, we have $%
S=S_{o}[\rho _{o},\theta _{o}]+S_{1}$, where
\begin{eqnarray}
S_{1} &=&\int \left[ i\hbar \rho _{1}\partial _{\tau }\theta _{1}+\frac{1}{2}%
\theta _{1}\cdot \hat{T}\cdot \theta _{1}+\frac{1}{2}\rho _{1}\cdot \hat{G}%
\cdot \rho _{1}\right]   \label{S1} \\
\theta _{1}\cdot \hat{T}\cdot \theta _{1} &=&-\theta _{1}{\bf \nabla }\cdot %
(\rho _{o}\left[ \hbar ^{2}/M\right] {\bf \nabla }\theta _{1})  \label{T} \\
\rho _{1}\cdot \hat{G}\cdot \rho _{1} &=&g_{d}\rho _{1}^{2}-(\hbar
^{2}/2M)[(\rho _{1}/\sqrt{\rho _{o}})\nabla ^{2}(\rho _{1}/\sqrt{\rho _{o}}%
)-\left( \rho _{1}^{2}/\rho _{o}^{3/2}\nabla ^{2}\sqrt{\rho _{o}}\right) ]
\label{G}
\end{eqnarray}
One can also integrate out the density fluctuations and obtain an effective
action of phase fluctuations $\theta _{1}$,
\begin{equation}
S_{{\rm phase}}=\int \left[ \frac{1}{2}\hbar ^{2}(\partial _{\tau }\theta
_{1})\cdot \hat{G}^{-1}\cdot (\partial _{\tau }\theta _{1})+\frac{1}{2}%
\theta _{1}\cdot \hat{T}\cdot \theta _{1}\right] .  \label{Sphase}
\end{equation}

Quantum fluctuations are caused by zero point motions of the excitations.
The low-lying excitations are phonons, corresponding to density oscillations
inside the atomic cloud. At high excitation energy, the period of density
oscillations become comparable to cloud size, the excitations leak out of
the cloud and become single-particle like, and cannot be adequately
described by the phase action eq. (\ref{Sphase}). The equations of motion of
the collective modes in real time $t$ can be obtained by first analytically
continuing the action eq.(\ref{S1}) to real time $(\tau \rightarrow it)$ and
then finding the stationary conditions. They are $\hbar \partial _{t}\theta
_{1}=\hat{G}\rho _{1}$ and $-\hbar \partial _{t}\rho _{1}=\hat{T}\theta _{1}$%
. The wavefunctions and the frequencies of the collective modes (labelled by
a set of quantum numbers $\alpha \equiv \{\alpha _{i}\}$) are the
eigenfunctions and eigenvalues of the operator $\hat{G}\hat{T}$, i.e.
\begin{equation}
\hat{G}\hat{T}u_{\alpha }=\hbar ^{2}\Omega _{\alpha }^{2}u_{\alpha }.
\label{GT}
\end{equation}
The $\Omega =0$ mode (corresponding to a constant phase) is excluded from
the collective excitation spectrum because it is simply the ground state.
Since all other modes are orthogonal to the $\Omega =0$ mode, they have zero
spatial average $--$ a fact that shall be of use to us later.

In 1956, Penrose and Onsager\cite{Penrose} pointed out a unified description
of Bose-Einstein condensation (BEC) for both noninteracting and interacting
systems based on the general structure of the density matrix $W({\bf r},{\bf %
r^{\prime }})=<\psi ^{+}({\bf r})\psi ({\bf r^{\prime }})>=\sum_{\alpha
}\lambda _{\alpha }\nu _{\alpha }^{*}({\bf r})\nu _{\alpha }({\bf r^{\prime }%
})$ , where $\lambda _{\alpha }$ and $\nu _{\alpha }$ are the eigenvalues $%
\lambda _{\alpha }$ and normalized eigenfunctions of $W$. BEC is
characterized \cite{Penrose} by the fact that below a transition temperature
$T_{c}$, one of the eigenvalues (say $\lambda _{0}$) becomes order $N$
compared with all others,
\begin{equation}
W=\lambda _{0}\nu _{0}^{*}({\bf r})\nu _{0}({\bf r^{\prime }})+K({\bf r},%
{\bf r^{\prime }}),\,\,\,\,\,\lambda _{0}/\lambda _{i\neq 0}\sim N,
\label{Penrose}
\end{equation}
where $K$ denotes the sum of those eigenstates $\lambda _{i\neq 0}$ and (for
homogeneous systems) has the property that $K({\bf r},{\bf r^{\prime }}%
)\rightarrow 0$ as $|{\bf r}-{\bf r^{\prime }}|\rightarrow \infty $. The
first term of eq.(\ref{Penrose}), which represents condensation, is referred
to off-diagonal-long-range order\cite{Yang}. For trapped Bose gases,
however, both ${\bf r}$ and ${\bf r^{\prime }}$ must remain inside the
atomic cloud. The large distance behavior of $W$ is masked by finite size
effects. In spite of this, the appearance of a $macroscopic$ eigenvalue as
described in eq.(\ref{Penrose}) remains a proper characterization of BEC\cite
{comment}.

To demonstrate the existence of macroscopic off-diagonal order in trapped
Bose gases in all dimensions at $T=0$, we write
\begin{equation}
W({\bf r},{\bf r^{\prime }})=<\sqrt{\rho ({\bf r})\rho ({\bf r^{\prime }})}%
e^{-i[\theta ({\bf r})-\theta ({\bf r^{\prime }})]}>=\sqrt{\rho _{o}({\bf r}%
)\rho _{o}({\bf r^{\prime }})}\left( e^{-F({\bf r},{\bf r^{\prime }}%
)}+Y({\bf r}, {\bf r}') \right) ,  \label{W}
\end{equation}
where $F({\bf r},{\bf r^{\prime }})\equiv \frac{1}{2}<[\theta _{1}({\bf r}%
)-\theta _{1}({\bf r^{\prime }})]^{2}>$, and \begin{eqnarray}
Y({\bf r}, {\bf r}') &=&\frac{-i}{2}\left\langle \left( \frac{\rho
_{1}({\bf r})}{\rho _{0}(%
{\bf r})}+\frac{\rho _{1}({\bf r}^{\prime })}{\rho _{0}({\bf r}^{\prime })}%
\right) \left( \theta ({\bf r})-\theta ({\bf r^{\prime }})\right)
\right\rangle   \nonumber \\
&&+\frac{\left\langle \rho _{1}({\bf r})\rho _{1}({\bf r}^{\prime
})\right\rangle }{4\rho _{o}({\bf r})\rho _{o}({\bf r^{\prime }})}-\frac{%
\left\langle \rho _{1}^{2}({\bf r})\right\rangle }{8\rho _{o}^{2}({\bf r})}-%
\frac{\left\langle \rho _{1}^{2}({\bf r}^{\prime })\right\rangle }{8\rho
_{o}^{2}({\bf r}^{\prime })}+O\left( \left( \frac{\rho _{1}}{\rho _{0}}%
\right) ^{3/2}\right)   \label{etc}
\end{eqnarray}
As we shall see later, the term $Y({\bf r}, {\bf r}') $ is smaller that the
first term in
eq.(\ref{W}) by at least a factor of $(\sigma /R)^{4}$
where $R$ is the size of the cloud. We can then ignore the term $Y({\bf r},
{\bf r}') $ in eq.(\ref{W}).
Using eq.(\ref{Sphase}), it is straightforward to show that
\begin{equation}
F({\bf r},{\bf r^{\prime }})=\sum_{\alpha }\left| <{\bf r}|\hat{G}%
^{1/2}|\alpha >-<{\bf r^{\prime }}|\hat{G}^{1/2}|\alpha >\right| ^{2}\frac{1%
}{2\hbar \Omega _{\alpha }}.  \label{F}
\end{equation}
where $<{\bf r}|\alpha >=u_{\alpha }({\bf r})$. At first sight, the phase
fluctuation appears to decrease with the size of the system $R$ because $%
u_{\alpha }^{2}$ scales as $R^{-d}$. This, however, is not true because the $%
\alpha $-sum (which depends on the number of excitations contributing
significantly to the phase fluctuation) is also $R$ dependent. To evaluate $F
$, we need to study the collective modes in greater detail. For systems with
large number of particles, it is known that Thomas-Fermi approximation (TFA)
becomes accurate\cite{Baym}. TFA ignores the gradient term in eq. (\ref{GP}%
). It implies that
\begin{equation}
\rho _{o}({\bf r})=\frac{M\omega ^{2}R^{2}}{2g_{d}}\left( 1-\frac{r^{2}}{%
R^{2}}\right) \Theta (R^{2}-r^{2}),\,\,\,\,\,\,\,\mu \equiv \frac{1}{2}%
M\omega ^{2}R^{2},  \label{density}
\end{equation}
where $\Theta (x)=1$ (or $0$) of $x>0$ (or $<0$). $R$ is the size of the
system. The number constraint $N=\int \rho _{o}$ implies that the system
size $R$ is related to $N$ as
\begin{equation}
R/{\sigma }=\left[ g_{d}N(d+2)/(\hbar \omega c_{d}\sigma ^{d})\right]
^{1/(d+2)},  \label{Rsigma}
\end{equation}
where $c_{d}\sigma ^{d}$ is the volume of a $d$-dimensional sphere of radius
$\sigma $. The fact the $\rho _{o}\propto g_{d}^{-1}$ in TFA means that $%
\hat{T}\propto g_{d}^{-1}$. In addition, TFA also implies that the term $%
\left( (\hbar ^{2}/4M)\rho _{1}^{2}/\rho _{o}^{3/2}\nabla ^{2}\sqrt{\rho _{o}%
}\right) $in $\hat{G}$ (eq.(\ref{G})) is negligible compared to $g_{d}$.
Furthermore, for density oscillations with period small compared to cloud
size (phonon regime, see below), $(\hbar ^{2}/4M)$ $[(\rho _{1}/\sqrt{\rho
_{o}})\nabla ^{2}(\rho _{1}/\sqrt{\rho _{o}})]<<$ $g_{d}$, and can also be
ignored. As a result, $\hat{G}=g_{d},$ and the explicit $g$ dependence in
eq.(\ref{GT}) is cancelled out. For harmonic potentials, eq.(\ref{GT})
becomes\cite{Stringari}
\begin{equation}
\frac{1}{2}\omega ^{2}{\bf \nabla }([R^{2}-r^{2}]{\bf \nabla }u_{\alpha })=%
\Omega _{\alpha }^{2}u_{\alpha },  \label{Legendre}
\end{equation}

For $d=1$, eq.(\ref{Legendre}) is the Legendre equation. The normalized
eigenfunctions are $u_{\ell }(r)=(2R)^{-1/2}P_{\ell }(r/R)$, where $P_{\ell
}(x)$ are Legendre polynomials. The frequency of $u_{\ell }$ is $\Omega %
_{\ell }=(\omega /\sqrt{2})\sqrt{\ell (\ell +1)}$, $\ell =1,2,..$. In $d=2$,
it is straightforward to show that the eigenfunctions are of the form $%
u_{n,m}(r,\phi )=e^{-im\phi }\sum_{\ell =|m|}^{n}c_{\ell }(r/R)^{\ell }$
with $n=|m|+2p,p=0,1,2,..$. The corresponding frequencies are $\Omega %
_{n,m}=(\omega /\sqrt{2})\sqrt{n^{2}-m^{2}+2n}$. We have thus established
statement $({\bf II})$. The collective mode in $d=3$ has been derived by
Stringari\cite{Stringari}. Because of the gapped nature of the collective
mode, the phase fluctuation eq.(\ref{F}) reduces its zero temperature form
for $k_{B}T<\hbar \omega $. In other words, the existence of macroscopic
off-diagonal order at $T=0$ immediately implies its existence at
sufficiently low temperatures. We shall therefore from now on focus on the $%
T=0$ case.

{\em Dilute condition and phonon cutoff } : To evaluate eq.(\ref{F}) , it is
necessary to discuss a number of implicit constraints implied by the low
energy model (eq.(\ref{S})), as well as how the collective modes behave as
their energy increases. First of all, to stay in the dilute limit, the
density at the center of the trap must be less that the inverse volume of
the atomic cutoff, $\rho _{o}({\bf 0})<(c_{d}a_{c}^{d})^{-1}$. This limits
the size of the cloud to be smaller than $R_{m}$,
\begin{equation}
\left( \frac{R}{\sigma }\right) ^{2}<\frac{g_{d}}{\frac{1}{2}\hbar \omega
c_{d}a_{c}^{d}}\equiv \left( \frac{R_{m}}{\sigma }\right)
^{2}\,\,\,\,\,\,\,\,\,:\,\,{\rm dilute\,\,\,condition}.  \label{dilute}
\end{equation}
For a cloud of radius $R$ satisfying eq.(\ref{dilute}), the collective modes
will be labelled by $d$ quantum numbers $\alpha _{i}$, $i=1,..d$, denoting
the number of nodes in $d$ different curvilinear directions. The spatial
variations of this mode along a curvilinear direction is given by a averaged
wavevector $k_{i}\sim \alpha _{i}/R$. The corresponding gradient energy is $%
\hbar ^{2}k_{i}^{2}/2M$. As the energy of the collective mode increases
(i.e. increasing $\alpha _{i}$), $k_{i}$ will increase to the point where
the condition $(\hbar ^{2}/4M)$ $[(\rho _{1}/\sqrt{\rho _{o}})\nabla
^{2}(\rho _{1}/\sqrt{\rho _{o}})]<<$ $g_{d}\rho_{1}^{2}$ breakdowns.
This occurs when $\hbar ^{2}k_{i}^{2}/2M\sim g_{d}\rho ({\bf
0})=\frac{1}{2}M\omega ^{2}R^{2}$
, which is $k_{i}=R/\sigma ^{2}$, or
\begin{equation}
\alpha _{i}=(R/\sigma )^{2}\equiv \alpha _{ph}\,\,\,\,\,\,{\rm %
:\,\,Phonon\,\,cutoff}\,\,\,\,\,\,  \label{TFcut}
\end{equation}
Obviously, both $R_{m}$ and $\alpha _{ph}$ are only defined up to some
numerical factor. The collective modes with quantum numbers below $\alpha
_{ph}$ are described by eq.(\ref{Legendre}).
For frequencies $\Omega >\Omega _{\alpha _{ph}}$, the excitation eq.(\ref{GT}
) will not be phonon like. Instead, they leak out of the cloud and become
single-particle like. The action eq.(\ref{S1}) is no longer sufficient.
Eventually, the average wavevector of the collective mode will approach the
``high energy'' cutoff $k_{i}\sim 1/a_{c}$, corresponding to
\begin{equation}
\alpha _{i}\sim R/a_{sc}\equiv \alpha ^{*},\,\,\,\,\,\,:\,\,{\rm %
high\,\,energy}\,\,{\rm cutoff},  \label{hecut}
\end{equation}
in which case the low energy model eq.(\ref{S}) is no longer valid.
Excitations between the phonon cutoff $\alpha _{ph}$ and the high energy
cutoff $\alpha ^{*}$ will be referred to as ``high energy'' modes. (Clearly
we need to have $R/a_{c}>(R/\sigma )^{2}$, (i.e. $\sigma /a_{c}>R/\sigma $)
for the cutoffs in eq.(\ref{TFcut}) and eq.(\ref{hecut}) to be consistent.
For $\sigma /a_{c}\sim 10^{2}$, this condition is easily satisfied even for
large clouds with $R/\sigma \sim 20$.

Now we return to eq. (\ref{W}). We note that the first term in
eq.(\ref{etc}) is odd in $\theta $ and vanishes by symmetry. For the second
term in eq.(\ref{etc}), it is straightforward to show\cite{Popov} that its
contribution from a
given phonon mode of average momentum $k\sim 1/R$ is down from the
corresponding phase fluctuation by $\hbar ^{2}k^{2}/\left( Mg_{d}\rho
_{0}\right) \sim \left( \sigma /R\right) ^{4}\ll 1$ for large clouds. We
can therefore ignor the density fluctuation term $etc$ in eq.(\ref{W}) as
mentioned before. Writing
$F({\bf r},{\bf r^{\prime }})$$%
=(\sum_{\alpha _{i}<(R/\sigma )^{2}}$ $+\sum_{(R/\sigma )^{2}<\alpha
_{i}<R/a_{c}})$$(...)$ $\equiv $$F_{ph}+F_{high}$, where $(...)$ is the
summand of eq.(\ref{F}). The leading $R$ dependence comes from $F_{ph}$. The
physics of $F_{high}$ and high energy density fluctuations is analogous to
the similar high energy contributions in homogeneous systems, which are
known to be less important than the phonon contributions for large $\left|
{\bf r}-{\bf r}^{\prime }\right| $\cite{Popov}.
By noting that within the phonon regime, (i.e. for excitations with $
\alpha_{i}<\alpha_{ph}$),  the operator $\hat{G}$ reduces to the constant
$g_{d}$,
Eq.(\ref{F}) implies that
\begin{equation}
F_{ph}({\bf r},{\bf r^{\prime }})=\sum_{\alpha _{i}<(R/\sigma )^{2}}\frac{%
g_{d}}{2\hbar \Omega _{\alpha }}\left| u_{\alpha }({\bf r})-u_{\alpha }({\bf %
r^{\prime }})\right| ^{2}.  \label{Fac}
\end{equation}

Denoting the spatial average and fluctuation of a function $f({\bf r})$ as $%
\overline{f}$ and $\Delta (f)$, i.e. $f=\overline{f}+\Delta (f)$, we have
\begin{equation}
F_{ph}({\bf r},{\bf r^{\prime }})=\gamma _{d}+\eta _{d}({\bf r},{\bf %
r^{\prime }}),\,\,\,\,\,\gamma _{d}=\frac{g_{d}}{c_{d}R^{d}}\sum_{\alpha
<(R/\sigma )^{2}}\frac{1}{\hbar \Omega _{\alpha }},  \label{gammad}
\end{equation}
and $\eta _{d}({\bf r},{\bf r^{\prime }})$ $=(g_{d}/2\hbar \Omega _{\alpha
}) $ $\sum_{\alpha <(R/\sigma )^{2}}$ $[\Delta (|u_{\alpha }({\bf r}%
)|^{2}+|u_{\alpha }({\bf r^{\prime }})|^{2})$ $-(u_{\alpha }({\bf r}%
)^{*}u_{\alpha }({\bf r}^{\prime })+{\rm c.c.})]$.

To estimate $F_{ph}$, we note that there is a one-to-one correspondence
between the solutions of eq.(\ref{Legendre}) and those of the equation
\begin{equation}
-(\omega ^{2}R^{2}/2){\bf \nabla }^{2}u=\Omega ^{2}u,  \label{LL}
\end{equation}
as both of them are labelled by the same quantum numbers. Moreover, the
solutions of eq.(\ref{Legendre}) resemble those of eq.(\ref{LL}) as the
number of nodes increases so that the gradient term becomes more important.
We thus approximate collective mode frequencies and wavefunctions in eq.(\ref
{Fac}) by those of eq.({\ref{LL}). The normalized functions $u_{\alpha }(%
{\bf r})$ will be replaced normalized plane waves $u_{{\bf k}}({\bf r})=e^{i%
{\bf k}\cdot {\bf r}}/\sqrt{c_{d}R^{d}}$, where ${\bf k}$ is a $d$
dimensional wave-vector. $k_{i}R$ is the correspondence of $\alpha _{i}$ as
it counts the number of nodes along the $i$-th (linear) direction. The
frequency $\Omega _{\alpha }$ in eq.(\ref{Fac}) corresponds to $\omega (Rk)/%
\sqrt{2}$ in eq.(\ref{LL}). With this approximation, eq.(\ref{Fac}) becomes
\begin{equation}
F_{ph}({\bf r},{\bf r^{\prime }})\approx \frac{1}{c_{d}R^{d}}\frac{\sqrt{2}%
g_{d}}{\hbar \omega }\int_{1}^{(R/\sigma )^{2}}\frac{{\rm d}^{d}(kR)}{(2\pi
)^{d}}\left( \frac{1-{\rm cos}{\bf k}\cdot ({\bf r}-{\bf r^{\prime }})}{kR}%
\right) .  \label{newFph}
\end{equation}
The constant and the cosine term in the bracket in eq.(\ref{newFph})
correspond to the average and fluctuation term ($\gamma _{d}$ and $\eta _{d}$
in eq.(\ref{gammad}) respectively). The lower limit is due to the fact that
the lowest collective mode has an average wavelength $R$ (or wavevector $1/R$
). }

Eq.(\ref{newFph}) implies that
\begin{equation}
F_{ph}^{d=1}(r,r^{\prime })=\frac{\sqrt{2}%
}{\pi }\left( \frac{g_{1}}{2\sigma \hbar \omega }\right) \left( \frac{%
\sigma }{R}\right) \left[2{\rm ln}\left( \frac{R}{\sigma }\right) -{\rm Ci}%
\left( \frac{2R^{2}x}{\sigma ^{2}}\right)+{\rm Ci}(2x)\right],
\end{equation}
\begin{equation}
F_{ph}^{d=2}(%
{\bf r},{\bf r^{\prime }})=\frac{\sqrt{2}}{4\pi }\left( \frac{g_{2}}{%
\hbar \omega \pi \sigma ^{2}}\right) \left( \frac{\sigma }{R}\right) ^{2}
\left[\left( \frac{R}{\sigma }\right) ^{2}-1-\frac{\sqrt{\pi }}{2}\Gamma
(1/2)\left[J_{o}(a){\bf H}_{-1}(a)-
{\bf H}(a)J_{-1}(a)\right]_{a=1}^{a=xR^{2}/\sigma
^{2}}\right],
\end{equation}
\begin{equation}
 F_{ph}^{d=3}({\bf r},{\bf r^{\prime }})=\frac{\sqrt{2}}{2\pi ^{2}}\left(
\frac{\sqrt{2}g_{3}}{\hbar \omega (4\pi /3)\sigma ^{3}}\right) %
\left( \frac{\sigma }{R}\right) ^{3}[\frac{1}{2}\left( \frac{R}{\sigma }%
\right) ^{4}-\frac{1}{2}-\frac{1}{x}(J_{1}\left( \frac{{R^{2}}x}{\sigma
^{2}}\right) -J_{1}(x))],
\end{equation}
 where $x=|{\bf r}-{\bf r^{\prime }}|/R$, Ci is
the Cosine integral, $J_{o},J_{-1}$ are Bessel functions, $\Gamma $ is the
Gamma function, and ${\bf H}_{o},{\bf H}_{-1}$ are the Struve functions. The
first and second terms in these expressions are the constant $\gamma _{d}$
and the function $\eta _{d}$ in eq.(\ref{gammad}). It is easy to show that $%
\eta_{d} \rightarrow 0$ as $|{\bf r}-{\bf r}'| \rightarrow \infty $. Even for
$|{\bf r}-{\bf r}'| /\sigma \sim 2$, $\eta _{d}<<\gamma _{d}$ provided
$R/\sigma >5$.  In other words, for
large clouds, we have $F_{ph}\sim \gamma _{d}$ when $|{\bf r}-{\bf r^{\prime
}}|\sim R/2$. Eq.(\ref{W}) and (\ref{F}), and the fact that $F\sim F_{ph}$
implies statement ${\bf (I)}$. Note that although the cutoff $\alpha _{ph}$
in our derivation contains an uncertain numerical factor, it does not affect
the $R$ dependence of our results. Our results also leads to the
interesting conceptual point that for a true $1d$ system with a $\delta $
-function potential of finite strength, the mean field off-diagonal form of
density matrix becomes exact as $R\rightarrow \infty $.

The anomalous $R$ dependence of our results can be qualitatively understood
simply as follows. The density of trapped bosons is not arbitrary, but
increases with size $R$ according to the scaling relation $\rho \sim R^{2}.$
For a bulk homogenous system, the dimensionless coupling constant is $\frac{
mg_{d}}{\hbar ^{2}}\rho ^{\left( d-2\right) /2},$ and so increasing density
implies stronger (weaker) coupling in $3D$ ($1D).$ Indeed, the $R$
dependence of the condensate depletion in $d-$dimension is obtained if we
replace the trapped boson system with a homogenous system of size $R$ and
density $\rho \sim R^{2}.$

{\em Atomic gases in quasi $1d$ and $2d$ limit :} Our previous discussions
show that atomic Bose condensates
exist in these quasi low dimensional regimes at sufficiently low (non-zero)
temperatures. These condensates can be detected through their collective
modes in Statement ${\bf (II)}$. In addition, the momentum distribution $n(k)
$ (i.e. Fourier transform of $W$) will reflect the size dependence of the
quantum fluctuation.

When only the lowest harmonic quantum state in the tightly confining
direction is occupied, the coupling constant in the effective quasi $1d$
and $2d$ theory is given by
\begin{equation}
g_{1}=2\hbar \omega _{\perp }a_{sc},\,\,\,\,\,\,g_{2}=2\sqrt{2\pi }\hbar
\omega _{z}a_{z}a_{sc},  \label{g1g2}
\end{equation}
where $\sigma _{z}$ and $\sigma _{\perp }$ are the widths of the ground
state harmonic oscillator along $z$ and in the $xy$-plane, and $a_{sc}$ is
still the s-wave scattering length in three dimension. The quantum
suppression parameters $\gamma _{d}$ read from first terms of $F_{ph}^{d}(%
{\bf r},{\bf r^{\prime }})$, $d=1,2,3$, are then
\begin{equation}
\gamma _{3}=\frac{3\sqrt{2}}{8\pi ^{2}}\left( \frac{a_{sc}}{\sigma }\right)
\left( \frac{R}{\sigma }\right) ,\,\,\,\,\,\gamma _{2}=\frac{1}{\sqrt{\pi %
^{3}}}\left( \frac{a_{sc}}{\sigma _{\perp }}\right) \sqrt{\frac{\omega _{z}}{%
\omega _{\perp }}},\,\,\,\,\,\gamma _{1}=\frac{2\sqrt{2}}{\pi }\left( \frac{%
\omega _{\perp }}{\omega _{z}}\right) \left( \frac{a_{sc}}{R}\right) {\rm ln}%
\left( \frac{R}{\sigma _{z}}\right) .  \label{last}
\end{equation}
\newline
From eq. (\ref{Rsigma}), we see that $\gamma _{d}\sim \left( a_{sc}\right)
^{(d+1)/(d+2)},$ and so these effects are particularly noticeable in systems
with large $a_{sc}.$

In summary, we have shown that the quantum fluctuations in trapped boson
systems exhibit anomalous dimensionality dependence. Our results may be
observeable in experiments on quasi-$1D$ and $2D$ systems with anisotropic
traps.

\end{document}